\documentstyle[sprocl,epsf,cite]{article}

\def\Journal#1#2#3#4{{#1} {\bf #2}, #3 (#4)}

\def\NPB{{\em Nucl. Phys.} B}
\def\PLB{{\em Phys. Lett.}  B}
\def\PRL{\em Phys. Rev. Lett.}
\def\PRD{{\em Phys. Rev.} D}

\def\PR{\em Phys. Rep.}

\def\NAT{\em Nature}
\def\APP{\em Astropart. Phys.}

\def\be{\begin{equation}}
\def\ee{\end{equation}}
\def\bea{\begin{eqnarray}}
\def\eea{\end{eqnarray}}

\newcommand{\gev}{\,{\rm GeV}}

\newcommand{\lsim}{
  \raisebox{0.2em}{$<$} \hspace{-0.75em} \raisebox{-0.2em}{$\sim$} }

\begin{document}

\title{Precise calculation of neutralino relic density in the minimal
supergravity model\footnote{This talk is based on the work \cite{NRR}.}}

\author{Takeshi Nihei}

\address{Department of Physics, Lancaster University, Lancaster LA1 4YB, England, UK} 


\maketitle\abstracts{We study precise calculation
of neutralino relic density in the minimal supergravity model. 
We compare the exact formula for the thermal average of the neutralino 
annihilation cross section times relative velocity with 
the expansion formula in terms of the temperature, including
all the contributions to neutralino annihilation 
cross section. 
We confirm that the expansion formula fails badly near s-channel poles. 
We show that the expansion method causes 5-10 \% error even far away
from the poles.}
%
%
\section{Introduction}

In recent years, observations of cosmic microwave background radiation 
have been providing experimental values for cosmological parameters
with some accuracy. 
Current experimental bound on the fractional energy density of cold dark 
matter is \cite{Boomerang,Maxima}
\begin{eqnarray}
\Omega_{\rm CDM}h^2 & = & 0.1 - 0.3,
\label{eqn:omegah^2}
\end{eqnarray}
where $h$ $\approx$ 0.7 is the parameter in the Hubble constant 
$H_0$ $=$ 100 $h$ km/sec/Mpc \cite{Hubble}. 
This bound (\ref{eqn:omegah^2}) is expected to be improved in the near 
future, so precise calculation of the relic density becomes important. 

Particle physics should provide a particle-theoretical explanation
for dark matter. 
The minimal supergravity model \cite{SUGRA} is one of the most promising 
candidate for new physics beyond the standard model. 
In this model, a discrete symmetry, so-called R-parity, is introduced
to avoid rapid proton decay. One of the consequences of introducing 
the R-parity is that the lightest superparticle (LSP) is stable, 
and the LSP becomes a good candidate for cold dark matter 
\cite{neutralino-dm}. 
The LSP in this model is mostly the lightest neutralino $\chi$ which is 
a mass eigenstate given by a linear combination of neutral gauginos 
and Higgsinos
\begin{eqnarray}
\chi & = & N_{11} \tilde{B} + N_{12} \tilde{W}^3 
+ N_{13} \tilde{H}^0_1 + N_{14} \tilde{H}^0_2. 
\label{eqn:chi-1}
\end{eqnarray}

In general, supersymmetric models have a huge number of unknown 
parameters. However, the minimal supergravity model has only a small
number of parameters, so it has reasonable predictive power. 
There are five unknown parameters in this model
\begin{eqnarray}
m_0, \ m_{1/2}, \ A, \ \tan \beta, \ {\rm sgn}(\mu),
\label{eqn:5-parameters}
\end{eqnarray}
where $m_0$, $m_{1/2}$ and $A$ represent a common scalar mass, a common 
gaugino mass, and a common trilinear scalar coupling, respectively. 
They parametrize soft supersymmetry breaking terms at the GUT scale
$\approx$ $2 \times 10^{16} \gev$. 
$\tan \beta$ is the ratio of vacuum expectation values of 
the two neutral Higgs fields. 
$\mu$ denotes the Higgs mixing mass parameter.
In the case of the minimal supergravity model, the absolute value of $\mu$ 
is determined from the condition for consistent radiative electroweak
symmetry breaking, and only the sign is a free parameter. 
In this work, we assume that there are no CP violating phases in these 
parameters (\ref{eqn:5-parameters}) for simplicity. 

%
%
\section{Calculation of the relic density}
In this section, we first briefly review the standard calculation of 
the neutralino relic density \cite{Kolb-Turner&Jungman-etal} 
in the minimal supergravity model. 
After that, we explain what is necessary to calculate it precisely. 

We evaluate the relic density at present, starting from thermal
equilibrium in the early universe. 
The time evolution of the neutralino number density 
in the expanding universe is described by the Boltzmann equation
\begin{eqnarray}
\frac{d n_\chi}{dt} + 3 H n_\chi & = & 
- \langle\sigma v_{\rm M\o l}\rangle 
\left(n_\chi^2 - n_\chi^{\rm eq 2}\right), 
\label{eqn:Boltzmann-eq}
\end{eqnarray}
where $H$ is the Hubble expansion rate. 
$n_\chi$ describes the actual number density of the neutralino, while 
$n_\chi^{\rm eq}$ is the number density which the neutralino would have
in thermal equilibrium. $\sigma$ denotes the cross section of the neutralino
pair annihilation into ordinary particles. 
$v_{\rm M\o l}$ is so-called M$\o$ller velocity which can be identified 
as the relative velocity between the two neutralinos. 
$\langle\sigma v_{\rm M\o l}\rangle$ represents the thermal average of 
$\sigma v_{\rm M\o l}$. 

In the early universe, the neutralino is in thermal equilibrium 
$n_\chi$ $=$ $n_\chi^{\rm eq}$. As the universe expands, the neutralino
annihilation process freezes out, and after that the number of the 
neutralinos in a comoving volume remains constant. 

In the annihilation cross section $\sigma$ in eq.~(\ref{eqn:Boltzmann-eq}), 
there are a lot of final states which 
contribute: $\chi \chi$ $\longrightarrow$ $f \bar{f}$,
$hh$, $WW$, $ZZ$, $Zh$, etc. Among these final states, fermion pairs
$f \bar{f}$ usually give dominant contributions. 

Using an approximate solution to the Boltzmann equation 
(\ref{eqn:Boltzmann-eq}), the relic density $\rho_\chi$ $=$ $m_\chi n_\chi$
at present is given by
\begin{eqnarray}
\rho_\chi & = & 1.66 \frac{1}{M_{\rm Pl}} 
\left(\frac{T_\chi}{T_\gamma}\right)^3 T_\gamma^3 \sqrt{g_*}
\frac{1}{ \int_0^{x_F}dx \langle\sigma v_{\rm M\o l}\rangle },
\label{eqn:relic-density}
\end{eqnarray}
where $x$ $=$ $T/m_\chi$ is a temperature of the neutralino normalized 
by its mass. $T_\chi$ and $T_\gamma$ are the present temperatures  
of the neutralino and the photon, respectively. 
The suppression factor $(T_\chi/T_\gamma)^3$ $\approx$ $1/20$ follows 
from the entropy conservation in a comoving volume \cite{reheating_factor}. 
$M_{\rm Pl}$ denotes the Planck mass. 
$x_F$ $\approx$ $1/20$ is the value of $x$ at freeze-out, and 
is obtained by solving the following equation iteratively:
\begin{eqnarray}
x_F^{-1} & = & \ln \left( \frac{m_\chi}{2 \pi^3} \sqrt{\frac{45}{2g_* G_N}}
\langle\sigma v_{\rm M\o l}\rangle_{x_F} x_F^{1/2} \right),
\label{eqn:freeze-out-temperature}
\end{eqnarray}
where $G_N$ is the Newton's constant, and $g_*$ represents the effective
number of degrees of freedom at freeze-out.

The relativistic formula \cite{SWO} for thermal average in 
eq.~(\ref{eqn:Boltzmann-eq}) can be written as an integration 
over one variable \cite{Gondolo-Gelmini}
\begin{eqnarray}
\langle\sigma v_{\rm M\o l}\rangle & = & 
\frac{1}{8 m_\chi^4 T K_2^2(m_\chi/T)} 
\int_{4 m_\chi^2}^\infty ds \, \sigma(s) (s-4m_\chi^2)\sqrt{s}
K_1\left(\frac{\sqrt{s}}{T}\right),
\label{eqn:thermal-average}
\end{eqnarray}
where $K_i$ ($i$ $=$ 1,2) are the modified Bessel functions. 
The cross section $\sigma(s)$ is a complicated function
of $s$ in general, so we have to evaluate the above integration 
numerically to calculate the thermal average. 

If the normalized temperature $x$ is small enough, we may be able to 
use an expansion formula in terms of $x$  for the thermal average 
(\ref{eqn:thermal-average}) 
\begin{eqnarray}
\langle\sigma v_{\rm M\o l}\rangle & = & a + bx. 
\label{eqn:expansion}
\end{eqnarray}
This expansion, neglecting the higher order of $x$, is widely used in 
literatures.
In the case that the neutralino is the LSP, the $a$ coefficient for 
a fermion pair production is proportional to the fermion mass due to 
the Majorana nature of the neutralino. 
On the other hand the $b$ coefficient for the same final state includes 
a contribution which is not proportional to the fermion mass, 
so the $b$ coefficient is much larger than the corresponding $a$ coefficient 
for each fermionic final state: $a$ $\ll$ $b$ \cite{Goldberg}.

In order to calculate the relic density precisely enough, we have to 
take into account the following items:
\begin{itemize}
\item[(i)] all the contributions to the annihilation cross section
\item[(ii)] the exact formula for the thermal average
\item[(iii)] solving the Boltzmann equation exactly
\item[(iv)] coannihilation (This becomes important when the mass of 
the next LSP is close to the LSP mass 
\cite{Griest-Seckel,Mizuta-Yamaguchi,ino-coan,stau-coan}.)
\end{itemize}

In this work, however, we concentrate on (i) and (ii) to compare the 
expansion formula with the exact one, and we don't take into account
(iii) and (iv) for simplicity. Namely, we use the approximate solution 
(\ref{eqn:relic-density}) to the Boltzmann equation, and we neglect 
coannihilation effects. 

We have derived analytic expressions for the exact annihilation cross 
section and have obtained $a$ and $b$ coefficients analytically for 
every contribution including interference terms. 
These interference terms are neglected in most literatures, 
but we found that they can give significant contributions in some cases.
Some of the analytic expressions can be found in literatures for both
the exact cross section \cite{Lopez-Nano-Yuan,Griest} and the expansion 
coefficients \cite{Eli-Ros-Lal,Drees-Nojiri}. 

%
%
\section{Thermal average --- exact formula vs. expansion}
Let us see the rough behavior of 
the integrand in eq.~(\ref{eqn:thermal-average}). In the case of 
interest, the argument of the function $K_1$ is much larger than
unity, since $T$ $\lsim$ $m_\chi/20$ and $\sqrt{s}$ $\geq$ $2 m_\chi$. 
Therefore the thermal average can be written as a convolution of the cross
section with a function which decays exponentially:
\begin{eqnarray}
\langle\sigma v_{\rm M\o l}\rangle & \approx & 
\int_{4 m_\chi^2}^\infty ds \, \sigma(s) F(s),
\label{eqn:convolution}
\end{eqnarray}
where the function $F(s)$ has an exponential suppression factor as
\begin{eqnarray}
F(s) & \propto & e^{-\sqrt{s}/T}. 
\label{eqn:rough-behavior}
\end{eqnarray}
Naively, it seems that the expansion should converge quickly, since the 
function $F(s)$ in eq.~(\ref{eqn:rough-behavior}) decays exponentially 
as $s$ increases \cite{SWO}. However this is not always true. 
It is known that the expansion method fails badly when the 
annihilation cross section changes rapidly 
with $s$ \cite{Gondolo-Gelmini,Griest-Seckel,Lopez-Nano-Yuan}. 
This happens, e.g, near s-channel poles and thresholds of new channels.
In the following we compare the expansion result with the exact one,
including all the contributions to the annihilation cross section. 

%
%
\section{Numerical results}
The spectrum and the couplings in the minimal supergravity model at the weak 
scale are calculated by solving renormalization group equations 
with boundary conditions at the GUT scale and implementing
radiative electroweak symmetry breaking. In this analysis, we use an
existing code ``SUSPECT'' \cite{suspect} to calculate the spectrum 
and the couplings. 

%
\begin{figure}[t]
\hspace*{-0.5cm}
\unitlength 1mm
\epsfxsize=12.0cm
\leavevmode\epsffile{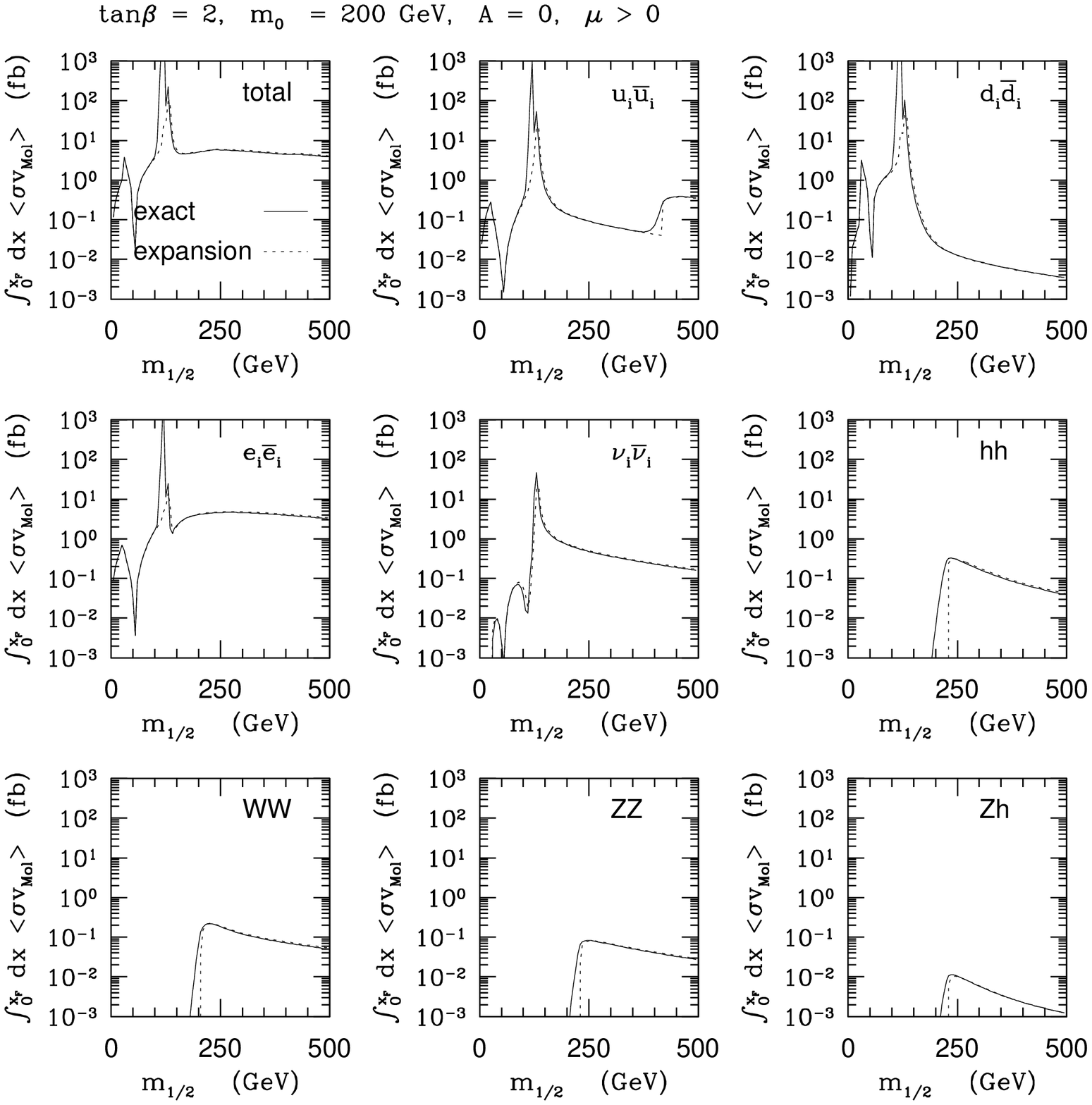}
\caption[fig1]{Integration of the thermal average $\int_0^{x_F}dx$ $ \langle\sigma v_{\rm M\o l}\rangle$ as a function of $m_{1/2}$ for $\tan \beta$ $=$ 2, 
$m_0$ $=$ $200 \gev$, $A$ $=$ 0, $\mu$ $>$ 0.}
\label{fig:intsv02+}
\end{figure}
%
%
In the following, we present the results of our calculation. 
Fig.~\ref{fig:intsv02+} shows the integration 
of the thermal average 
$\int_0^{x_F}dx$ $ \langle\sigma v_{\rm M\o l}\rangle$ 
as a function of $m_{1/2}$ for $\tan \beta$ $=$ 2, 
$m_0$ $=$ $200 \gev$, $A$ $=$ 0, $\mu$ $>$ 0. 
Note that the relic density in eq.~(\ref{eqn:relic-density}) is 
proportional to the inverse of this integration. 
The solid line and the dashed line represent the exact result and the 
expansion result, respectively.

We have ploted the total contributions and the individual contributions
for relevant final states. 
The exact results show similar behaviors with those in a previous 
analysis \cite{Baer-Brhlik}. 

There are two peaks in the total contributions. One is from the $Z$ boson 
contribution, and it takes place when $2 m_\chi$ $\approx$ $m_Z$ is 
satisfied.
The other is from the lightest Higgs ($h$) contribution, and it happens 
when $2 m_\chi$ $\approx$ $m_h$.
Around the poles, we see huge difference between the exact result and 
the expansion. Away from the poles, the difference is not so huge. 

\begin{figure}[t]
\vspace*{-5cm}
\hspace*{0.5cm}
\unitlength 1mm
\epsfxsize=10.0cm
\leavevmode\epsffile{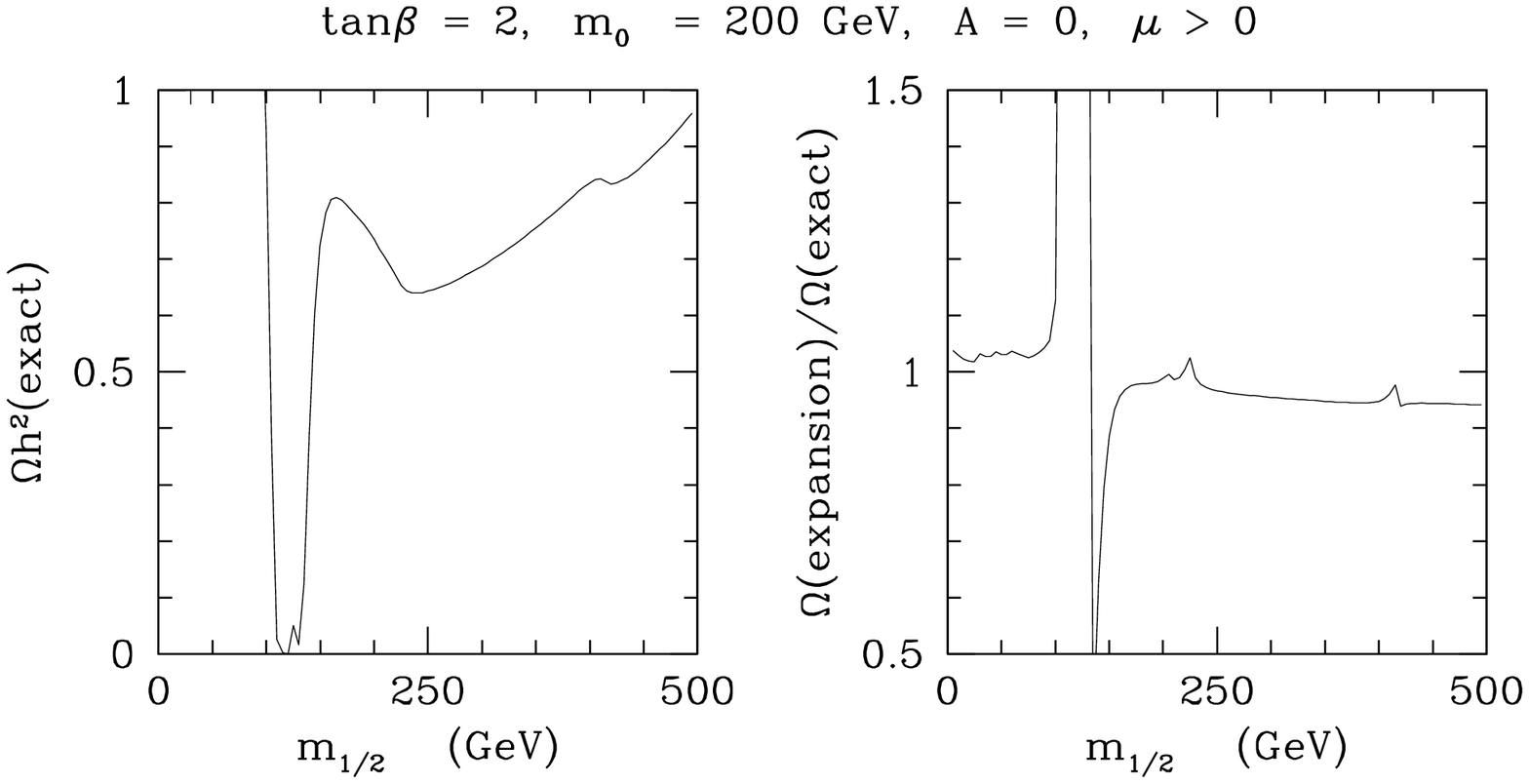} \\
\hspace*{3.05cm} (a) \hspace{4.3cm} (b)
\caption[fig2]{The relic density $\Omega h^2$ (a) and the ratio $\Omega_{\rm expansion}/\Omega_{\rm exact}$ (b) for $\tan \beta$ $=$ 2, $m_0$ $=$ $200 \gev$, $A$ $=$ 0, $\mu$ $>$ 0.}
\label{fig:omega02+}
\end{figure}

We plot the relic density vs. $m_{1/2}$ in Fig.~\ref{fig:omega02+} (a) 
for the same parameters as those in Fig.~\ref{fig:intsv02+}. 
In the region of the poles, we see a dip
of the relic density. Away from the poles, the relic density is too 
large to satisfy the experimental constraint (\ref{eqn:omegah^2}). 

%
%
%
\begin{figure}[t]
\vspace*{-5cm}
\hspace*{0.5cm}
\unitlength 1mm
\epsfxsize=10.0cm
\leavevmode\epsffile{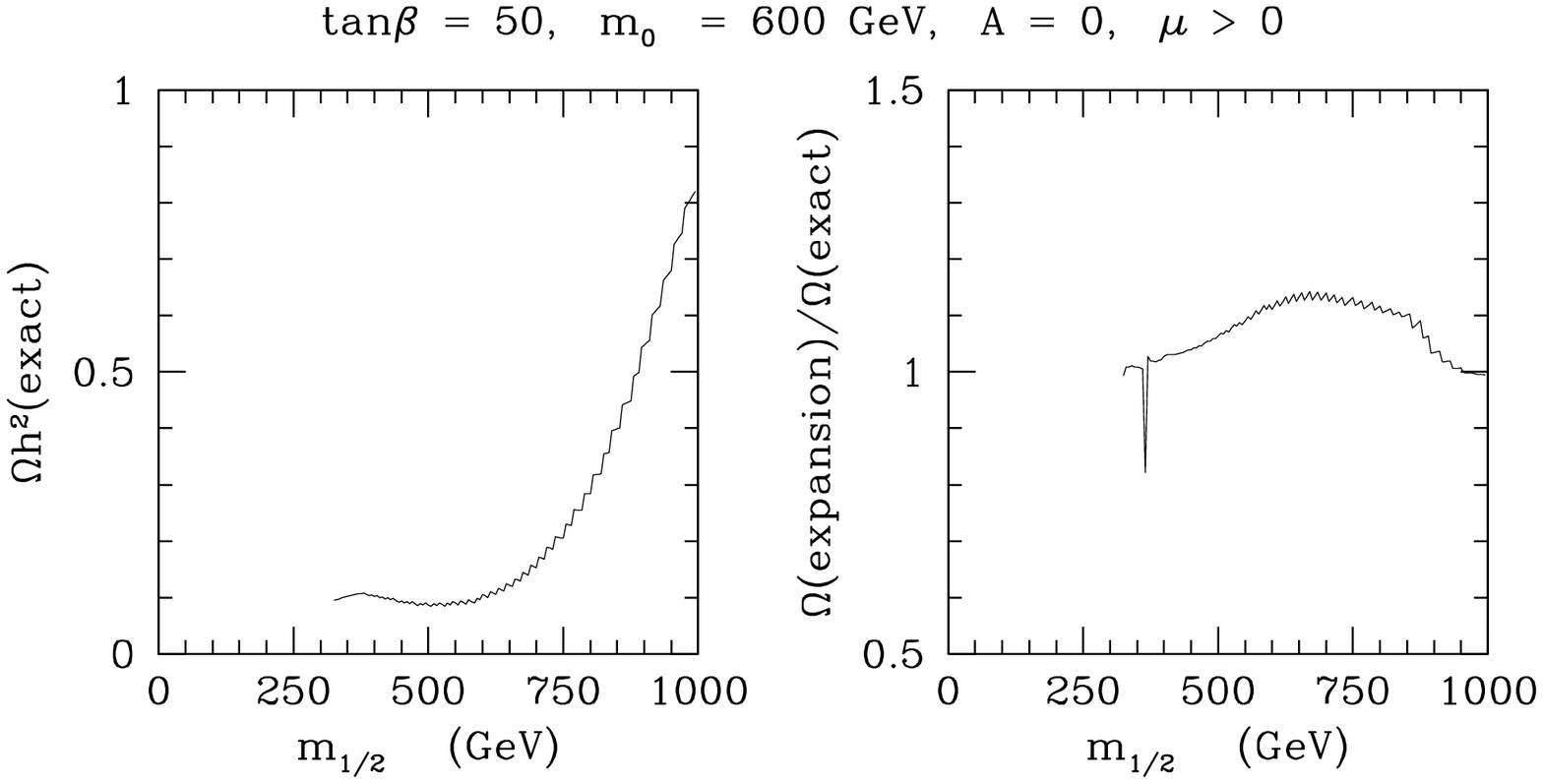} \\
\hspace*{3.05cm} (a) \hspace{4.3cm} (b)
\caption[fig3]{Similar results to Fig.~\ref{fig:omega02+} for $\tan \beta$ $=$ 50, $m_0$ $=$ $600 \gev$, $A$ $=$ 0, $\mu$ $>$ 0.}
\label{fig:omega50+}
\end{figure}
%

In Fig.~\ref{fig:omega02+} (b), 
we plot the ratio $\Omega_{\rm expansion}/\Omega_{\rm exact}$,
where $\Omega_{\rm exact}$ ($\Omega_{\rm expansion}$) denotes the relic 
density calculated with the exact (expansion) formula for the thermal 
average. 
In the pole regions, the expansion is quite different from the exact one
by a large factor \cite{Lopez-Nano-Yuan}. 
Also we find a difference of about 5 \% in this ratio even far away 
from the pole.

In a very large $\tan \beta$ case, the result has a different feature. 
In Fig.~\ref{fig:omega50+} (a) and (b),
we show similar results for $\tan \beta$ $=$ 50,
$m_0$ $=$ $600 \gev$, $A$ $=$ 0, $\mu$ $>$ 0. 
The relic density is small due to enhancements of 
the pseudoscalar Higgs exchange contribution and the heavier 
neutral Higgs exchange contribution in the $b \bar{b}$ final state
for a large $\tan \beta$. 
The are two reasons for these enhancements. First, couplings
of the pseudoscalar Higgs and the heavier neutral Higgs 
to the bottom quark become large for a large $\tan \beta$. Secondly, 
masses of heavy Higgses become smaller in a large $\tan \beta$ region. 
Because of these enhancements in the annihilation cross section, 
the relic density in Fig.~\ref{fig:omega50+} (a)
becomes small enough to satisfy 
the experimental constraint (\ref{eqn:omegah^2}) for
a wide region of $600 \gev$ $<$ $m_{1/2}$ $<$ $800 \gev$. 
As for the ratio of the expansion result to the exact one 
in Fig.~\ref{fig:omega50+} (b), 
we see a difference more than 10 \% for $600 \gev$ $<$ $m_{1/2}$ 
$<$ $800 \gev$. Note that we find this relatively large difference
in the interesting region where the relic density satisfies 
the experimental constraint (\ref{eqn:omegah^2}).

\begin{figure}[t]
\hspace*{3cm}
\unitlength 1mm
\epsfxsize=5.0cm
\leavevmode\epsffile{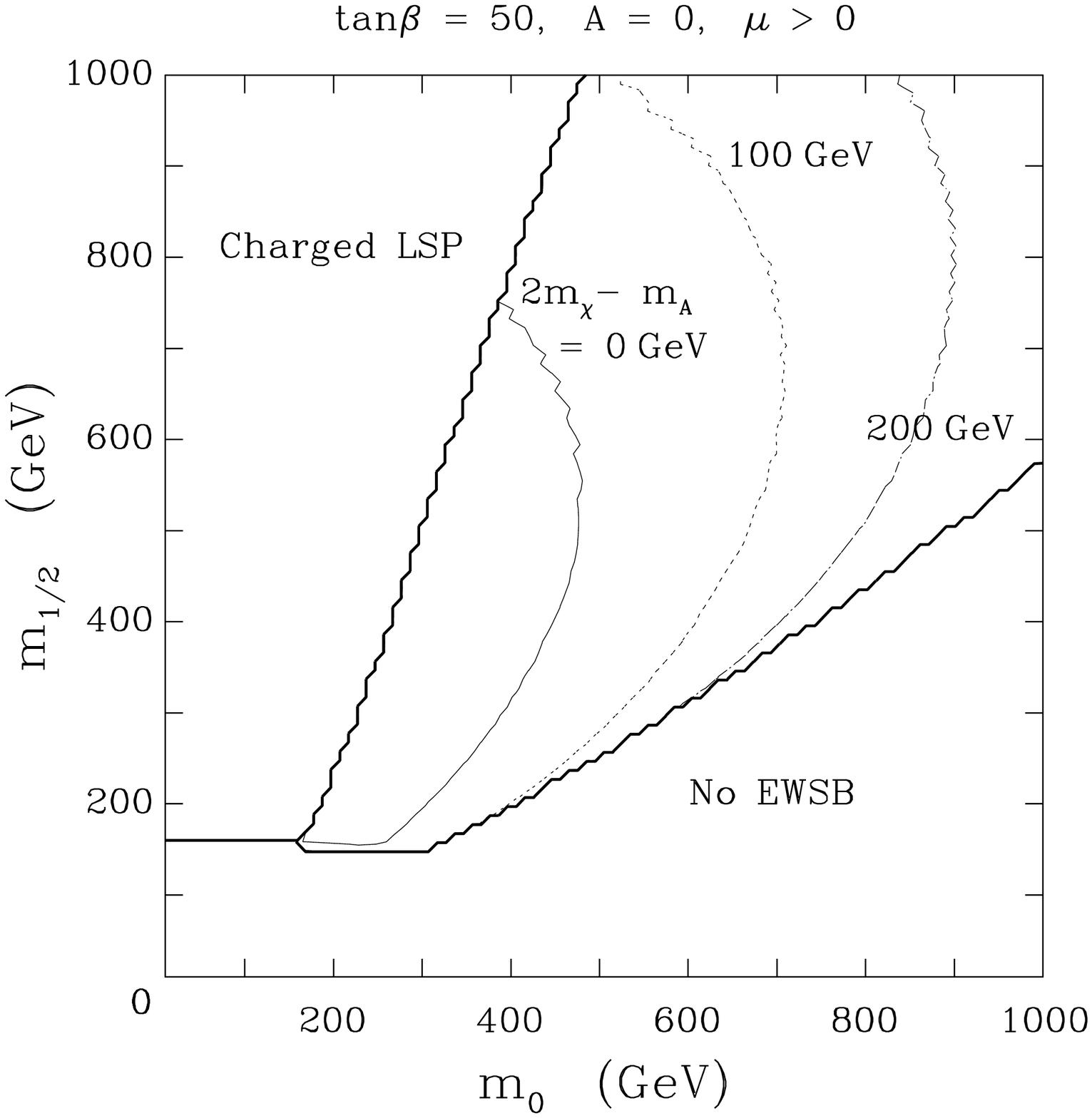}
\caption[fig4]{A contour plot of $2 m_\chi - m_A$ in the $m_0$--$m_{1/2}$ plane for $\tan \beta$ $=$ 50, $A$ $=$ 0 and $\mu$ $>$ 0. The region `Charged LSP' is excluded because the lighter stau is the LSP in this region. In the region `No EWSB', the electroweak symmetry breaking does not occur.}
\label{fig:pole}
\end{figure}

We show a contour plot for the quantity $2 m_\chi - m_A$ 
for $\tan \beta$ $=$ 50, $A$ $=$ 0 and $\mu$ $>$ 0 in 
Fig.~\ref{fig:pole}, where $m_A$ denotes the pseudoscalar Higgs mass. 
Along the thin solid line, this quantity vanishes so that 
there are enhancements in the annihilation cross section from the 
pseudoscalar Higgs pole contribution and the heavier neutral Higgs 
pole contribution through the $b \bar{b}$ final state. 
Fig.~\ref{fig:pole} shows that the region $600 \gev$ $<$ 
$m_{1/2}$ $<$ $800 \gev$ in Fig.~\ref{fig:omega50+} (b), 
where the deviation of $\Omega_{\rm expansion}$
from $\Omega_{\rm exact}$ is about 10 \%, differs from the pole 
region (i.e. the zero of  the quantity $2 m_\chi - m_A$) 
in $m_0$ by about $150 \gev$. Despite of this difference, we still
have a sizable error of 10 \% in the expansion result, as we have
seen in Fig.~\ref{fig:omega50+} (b).

%
%

In Fig.~\ref{fig:omega02+} and Fig.~\ref{fig:omega50+}, 
we find 5-10 \% difference even away from the pole. 
Thus the expansion never succeeds to approximate the exact result 
within the accuracy of 5 \%. 
This implies that the coefficient of the next order $cx^2$ in the 
expansion (\ref{eqn:expansion}) has the same order with the $b$ 
coefficient $c/b$ $\approx$ $O(1)$.

%
%
\section{Conclusions} 
We have presented results of precise calculation
of neutralino relic density in the minimal supergravity model. 
We have calculated all the contributions to neutralino annihilation 
cross section analytically. 
We have compared the exact formula for the thermal average 
of the annihilation cross section times relative velocity with
the expansion formula in terms of the temperature. 
We have confirmed that the expansion formula fails badly near 
s-channel poles. 
We have shown that the expansion method causes 5-10 \% error even 
far away from the poles.
%
%
\section*{Acknowledgments}
The author was supported in part by PPARC grant PPA/G/S/1998/00646. 
%
%
%


\section*{References}
\begin{thebibliography}{99}
%
%
\bibitem{NRR}
T. Nihei, L. Roszkowski and R. Ruiz, in preparation. 
%
%
\bibitem{Boomerang}
P. de Bernardis {\it et.al.},
\Journal{\NAT}{404}{955}{2000}.
%
\bibitem{Maxima}
A. Balbi {\it et.al.},
astro-ph/0005124. 
%
\bibitem{Hubble}
W. Freedman, \Journal{\PR}{333}{13}{2000}. 
%
%
\bibitem{SUGRA}
For reviews on the minimal supergravity model, see for instance, 
H.P. Nilles, \Journal{\PR}{110}{1}{1984};
P. Nath, R. Arnowitt and A.H. Chamseddine, 
Applied $N=1$ Supergravity, World Scientific, Singapore (1984). 
%
%
\bibitem{neutralino-dm}
J. Ellis, J.S. Hagelin, D.V. Nanopoulos, K.A. Olive and M. Srednicki, 
\Journal{\NPB}{238}{453}{1984}. 
%
%
\bibitem{Kolb-Turner&Jungman-etal}
For reviews on calculations of the relic density, see for instance, 
E.W. Kolb and M.S. Turner, {\em The Early Universe}, Addison-Wesley (1990);
G. Jungman, M. Kamionkowski and K. Griest, \Journal{\PR}{267}{195}{1996}.
%
%
\bibitem{reheating_factor}
K.A. Olive, D. Schramm and G. Steigman, 
\Journal{\NPB}{180}{497}{1981}. 
%
%
\bibitem{SWO}
M. Srednicki, R. Watkins and K.A. Olive, \Journal{\NPB}{310}{693}{1988}.
%
%
\bibitem{Gondolo-Gelmini}
P. Gondolo and G. Gelmini, \Journal{\NPB}{360}{145}{1991}.
%
%
\bibitem{Goldberg}
H. Goldberg, \Journal{\PRL}{50}{1419}{1983}.
%
%
\bibitem{Griest-Seckel}
K. Griest and D. Seckel, \Journal{\PRD}{43}{3191}{1991}.
%
\bibitem{Mizuta-Yamaguchi}
S. Mizuta and M. Yamaguchi, \Journal{\PLB}{298}{120}{1993}.
%
\bibitem{ino-coan}
J. Edsj\"o and P. Gondolo, \Journal{\PRD}{56}{1879}{1997}.
%
\bibitem{stau-coan}
J. Ellis, T. Falk, K.A. Olive and M. Srednicki, 
\Journal{\APP}{13}{181}{2000}.
%
%
\bibitem{Lopez-Nano-Yuan}
J.L. Lopez, D.V. Nanopoulos and K. Yuan, \Journal{\PRD}{48}{2766}{1993}.
%
%
\bibitem{Griest}
K. Griest, M. Kamionkowski, and M.S. Turner, \Journal{\PRD}{41}{3565}{1990}.
%
%
\bibitem{Eli-Ros-Lal}
J. Ellis, L. Roszkowski and Z. Lalak, \Journal{\PLB}{245}{545}{1990};
K.A. Olive and M. Srednicki, \Journal{\NPB}{355}{208}{1991}.
%
%
\bibitem{Drees-Nojiri}
M. Drees and M. Nojiri, \Journal{\PRD}{47}{376}{1993}.
%
%
\bibitem{suspect}
A. Djouadi, J.-L. Kneur and G. Moultaka, the code available on 
the web at 
http://www.lpm.univ-montp2.fr:7082/\~{\mbox{}}kneur/suspect.html. 
%
%
\bibitem{Baer-Brhlik}
H. Baer and M. Brhlik, \Journal{\PRD}{53}{597}{1996}.
%
%
\end{thebibliography}
\end{document}